# Open-Air, Broad-Bandwidth Trace-Gas Sensing with a Mid-Infrared Optical Frequency Comb

Lora Nugent-Glandorf[*], Fabrizio Giorgetta[§] and Scott A. Diddams[*]

[*]Time and Frequency Division, National Institute of Standards and Technology, 325 Broadway, Boulder, CO 80305, fax: 303-497-6461, lng@boulder.nist.gov, scott.diddams@nist.gov
[§]Optoelectronics Division, National Institute of Standards and Technology, 325 Broadway, Boulder, CO 80305

A mid-Infrared frequency comb is produced via an optical parametric oscillator (OPO) pumped by an amplified 100 MHz Yb:fiber mode-locked laser. We use this source to make measurements of the concentration of the atmospherically-relevant species of $CH_4$ and $H_2O$ over a bandwidth of 100 nm centered at 3.25 $\mu$m. Multiple absorption lines for each species are detected with millisecond acquisition time using a virtually-imaged phased array (VIPA) spectrometer. The measured wavelength-dependent absorption profile is compared to and fitted by a model, yielding quantitative values of the atmospheric concentration of both $CH_4$ and $H_2O$ in a controlled indoor environment, as well as over a 26 m open air outdoor path.

## 1. Introduction

Mid-infrared (MIR, 3-8 $\mu$m) laser spectroscopy [1] has been a topic of continued and growing interest over the past years, motivated in part by the strong absorptive features that constitute the unique spectroscopic "fingerprints" of a wide variety of molecular species important for applied and basic science [2]. Along these lines, specific application areas include trace gas sensing for environmental [3], energy [4] and medical diagnostics [5], earth sciences [6, 7] and recently time resolved spectroscopy [8]. However, the spectral region is also one of interest in the area of high precision spectroscopy for metrology, basic molecular physics, and precision measurements [9].

Many MIR spectroscopic applications benefit from broad bandwidth or continuous tunability in the laser source. For example, in the case of trace gas sensing, broad bandwidth provides for the possibility to simultaneously observe different compounds and discriminate one from the other. While thermal sources cover tremendous bandwidths, laser based MIR sources have the benefit of high brightness and a coherent beam, which enables propagation over long distances, cavity enhancement, microscopic modalities and coherent detection schemes. MIR laser-based sources in this category that have been employed for spectroscopy include broadly tunable quantum-cascade lasers [10, 11], supercontinuum sources [12], femtosecond mode-locked lasers [13], parametric oscillators [14, 15] and frequency comb sources [16, 17]. A particular advantage of the frequency comb is that it can simultaneously provide large bandwidth in a single measurement along with high spectral resolution and accuracy.

Within this context, direct spectroscopy with broadband femtosecond laser sources and frequency combs has been applied to various applications and measurements [4, 6, 8, 18-22]. In this paper we report on the use of a 3 micron broad bandwidth MIR frequency comb source in combination with a high-resolution Virtual-Image Phased Array (VIPA) spectrometer for trace gas detection of atmospheric gases over open-air paths of up to 26 m in length. A particular goal of this work is to explore the applicability and limits of the MIR source and VIPA detection for fast and accurate detection of methane and water. The system is described in detail and experiments are performed to quantify the system parameters, sensitivity, and precision in open path measurements both inside and outside the lab.

## 2. Experiment

Figure 1 outlines the generation of the coherent, mid-infrared light source, which is described in detail elsewhere [23]. The output of a Yb:fiber mode-locked femtosecond oscillator [24] (100 MHz repetition rate, 80 fs pulses) is temporally stretched then amplified in a double-clad Yb:fiber amplifier to produce 2 W of broad-bandwidth light centered at 1060 nm with a pulse width of 120 fs. This laser pumps an optical parametric oscillator (OPO) based on a 2 mm thick periodically-poled lithium niobate (PPLN) nonlinear crystal [15]. Through non-linear parametric down-conversion, the 1060 nm photons are split into signal and idler photons ($1/\lambda_{pump}=1/\lambda_{sig}+1/\lambda_{idler}$), where the signal field in the region of 1.4-1.8 µm is resonant in the OPO cavity, while the idler wavelengths are efficiently out-coupled. The OPO can be continuously tuned by changing the poling period of the fan-out PPLN crystal, accessing a wide range of mid-infrared wavelengths from 2.6-4.4 µm. In this study, we focus on the idler centered at 3.25 µm which covers the methane $\nu_3$ band. While full frequency stabilization of the comb is not necessary for atmospheric measurements due to the gigahertz linewidths arising from atmospheric pressure broadening, we utilize the intrinsic frequency comb structure of the optical spectrum to lock the length of the OPO cavity. As described in ref [15], spurious visible light leaking from the OPO from non-phase matched processes contains a heterodyne beat which corresponds to the difference of the offset frequencies of the signal and the idler combs (e.g. $2f_{0,idler} - f_{0,signal}$). Stabilization of this beat to an external radio frequency reference via a PZT-actuated mirror on one end of the OPO cavity effectively fixes the idler frequency comb spectrum relative to the pump, and creates a stable mid-IR spectrum over a sufficient period of time for our experiments (several hours without losing lock). Complete stabilization of the mid-IR frequency comb is possible [15], and could be useful for applications focused on higher-resolution spectroscopy.

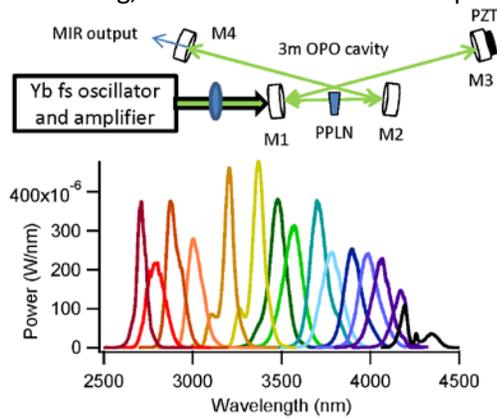

**Figure 1:** A 100 MHz, 120 fs, 2W Yb based oscillator and amplifier, centered at 1060 nm, is used to pump a linear OPO cavity containing a fanout PPLN non-linear crystal. The resonant cavity produces a signal and an idler, where the idler is coupled out at M4. The idler can be tuned by tuning the poling period of the PPLN crystal from 2.6-4.4 µm.

In the following, we study the feasibility of using this broad bandwidth mid-IR source to investigate real-time monitoring of methane and water concentrations in open-air and over long-path lengths. To this end, we test the stability and accuracy of our source and detection system with three different samples: a short 20 cm glass cell with Brewster windows, a long pass Herriott cell (26 m path length), and two open-air scenarios, one inside the lab and one in a true outdoor space (see Figure 2). We launch the mid-IR light through the sample and collect the data with a compact mid-IR spectrometer. The spectrometer collects spectral data in a 2-dimensional format by using a VIPA (Virtual-Image Phase Array) etalon that is cross-dispersed with a diffraction grating [23, 25]. After the light has exited the sample, the beam is focused into a mid-IR air-core photonic crystal fiber (PCF) [26]. This allows consistent alignment into the spectrometer when changing between different samples. The light from the fiber is collimated and then focused into the entrance window of the VIPA optic with a cylindrical lens. The VIPA is a tilted etalon (2 cm thick Silicon, ~50 GHz free spectral range), which gives large angular dispersion versus wavelength. The coupled beam is reflected back and forth between a highly reflective back surface (99.9 %) and a partially reflective front surface (98%) from which the dispersed spectral elements exit at varying angles. The output of the VIPA is then cross-dispersed with a standard diffraction grating blazed for 3 µm light (450 lines/mm), whose dispersion is chosen to separate the orders of the VIPA etalon into a 2-dimensional spectrum.  The spectrometer occupies approximately 30 cm by 60 cm, has a resolving power of ~150,000 and can capture ~80-100 nanometers of bandwidth (depending on imaging conditions) in a single spectrum acquired on milli- or sub-millisecond timescales.  To collect the dispersed optical spectrum after the sample, the spectrometer output is imaged onto a liquid-$N_2$ cooled InSb array camera (640 x 512 pixels). A 2.5-5 µm bandpass cold filter inside the camera minimizes excess pump and signal light, or stray background light, from hitting the camera. Figure 2 describes the spatial layout of each sample set-up. The exit beam of the 20 cm cell and the multi-pass Herriott cell (Figure 2 (a) and (b), respectively) is directly focused into the PCF with a $CaF_2$ lens. When the beam is sent into free space, as in the case of the in-lab path and the outdoor path, a retro-reflection mirror is used to return the beam to the VIPA spectrometer. The return beam is collected with a Cassegrain reflective telescope. As the beam path progresses in free space, the spatial divergence of the mid-IR beam results in a large beam profile (~5 cm diameter). The telescope captures the large beam and reduces it to a manageable size for focusing into the PCF fiber. Another advantage of the telescope is that small variations in alignment of the free space beam are minimized through the telescope imaging system.

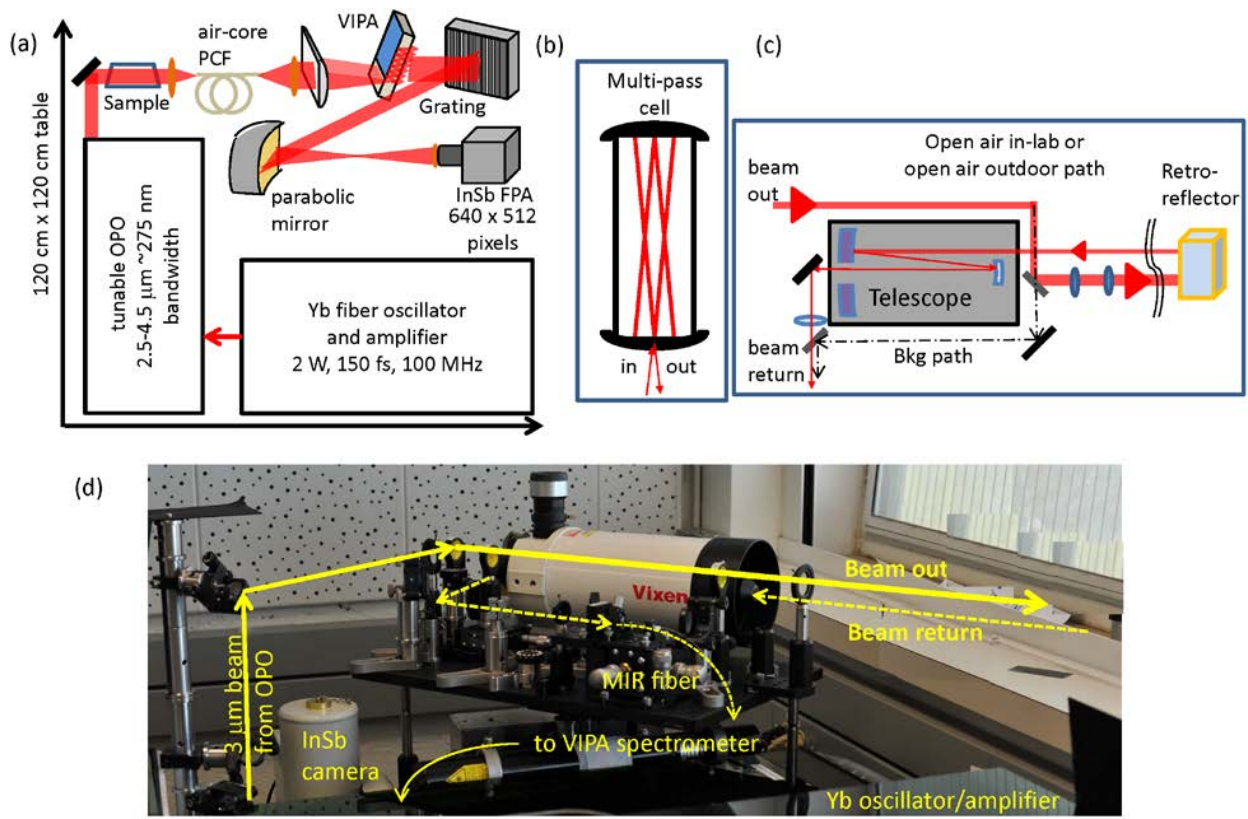

**Figure 2: (a)** The physical set up of the 120 cm x 120 cm mobile laser table housing the Yb oscillator/amplifier, the OPO, and the VIPA spectrometer. **(b)** In the place of the sample in (a), a multi-pass Herriott cell was used to determine the sensitivity of the system.     **(c)** The layout of the output beam to sample the air, the collection telescope and optics.  The telescope and optics sit above the laser and spectrometer on the 120 cm x 120 cm table as can be seen in **(d).**  The retro-reflection mirror is outside the laboratory, at a distance of ~13 m.  The beam passes through an IR transparent window.

## 2.1. Analyzing VIPA images

Once the two-dimensional images from the VIPA spectrometer are collected, they are analyzed as follows. For each 2D absorption spectrum, three images need to be acquired. First, a dark image is taken, representing the background noise of the spectrometer. This is accomplished with the beam blocked far away from the camera in order to eliminate laser light, but accommodate any residual background light. Secondly, a background image is acquired. For the sample cells (samples shown in Fig. 2 (a) and (b)), this requires pumping out of the cell with a small turbo pump to approximately $1\times10^{-5}$ Torr. For the long-path free space beam, a short free space path (shown as "bkg path" in Fig. 2 (c)) is set up with flip mirrors to send the beam directly into the telescope. The reported path length distances for absorption are the difference of the long-path distance image and the short path distance image. Once the dark and the background images are recorded, the sample images can be acquired. While this is the most ideal and rigorous approach to background correction, we later determined that only the sample image was needed and the background (including the dark image) could be removed by fitting the background of 0% transmission around a peak of interest. This method is discussed later in section 3.2.

An example of a series of images representing the rigorous background subtraction is shown in Figure 3. Fig. 3(a) is a raw data image of the 20 cm sample cell with 2 Torr pure methane. This allows for determination of the free spectral range of the VIPA etalon. Along the vertical range (VIPA dispersion direction) of the 640 x 512 pixel array of the InSb camera, the spectrum repeats itself, as indicated in Fig. 3(a) with the yellow circles identifying a single repeated absorption feature. In order to correctly display the spectrum in a traditional linear format, the free spectral range must be known. As explained in a previous publication [23], the resolution of the VIPA spectrometer (~700 MHz at this wavelength and grating angle) does not permit the 100 MHz comb lines to be resolved, so the images appear as stripes in the VIPA resolution direction. To lineate the spectrum, a grid of points is created via an automated computer program. The height of the grid in the vertical direction is determined by the free spectral range with 5 total points along the stripe. The horizontal grid is determined by the centers of the stripes in the grating direction (Fig. 3(b)). The grid is necessary due to tilt and curvature of the spectral stripes (see zoomed in portion of Fig. 3(d)) that arise from the properties of the VIPA spectrometer and the imaging system itself [27].

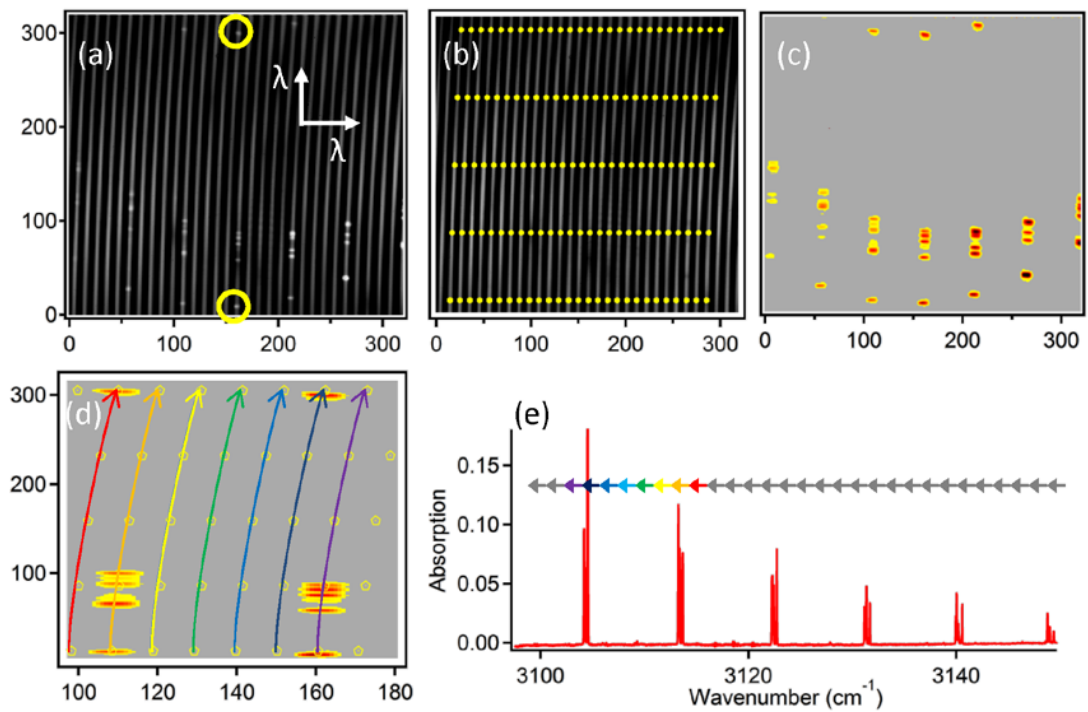

**Figure 3: (a)** A sample image of a 20 cm cell with 2 Torr methane. The free spectral range (FSR) of the VIPA is easily seen as a repeating pattern in the absorption features. The circles indicate an identical absorption feature. **(b)** A 2-dimensional grid is created on the VIPA lines from a background only image (no methane) based on the FSR determined from (a). **(c)** The methane lines can be clearly seen by dividing signal image (a) by background image (b). **(d)** A zoom in of (c) showing how each VIPA stripe is fit with a polynomial. At each point on the curve, 5 pixels horizontal to the VIPA stripe are averaged. **(e)** The stripes are concatenated to create a typical lineout absorption plot.

First, the no-light image is subtracted from each of the signal and background images. Secondly, for each VIPA stripe in the signal and background image (Fig. 3 (a)-(b), respectively), a parabolic function is fitted to the five points on the grid (Fig. 3(d)). Next, the data analysis program steps through each vertical pixel along the stripe and sums the counts from 5 horizontal pixels around the center pixel. This is done for each consecutive stripe, and the lineout spectrum is concatenated as shown by the coordinated color arrows in Fig. 3(e). The final lineout absorption spectrum is calculated as follows (shown in Fig. 3(e)):

$$\text{Absorbance} = -ln\left[L_S/L_B\right], \qquad \text{Equation 1}$$

where $L_S$ is the line spectrum of the sample image (after subtracting the dark image), and $L_B$ is the background image (after subtracting the same dark image). After obtaining the lineout spectrum, knowledge of the resolution of the spectrometer is necessary for proper fitting of the absorption data in order to determine concentrations of chemical species. Ref [23] describes in detail the resolution of the 2-D VIPA spectrometer (700 MHz, mentioned above) as well as the absorption sensitivity of the system ($4.5 \times 10^{-4}$ absorbance units). Figure 4 demonstrates the speed at which broad-bandwidth high-resolution data can be taken. In this experiment, we allowed methane to flow into the 20 cm cell and recorded the absorption at 375 frames/s with a reduced 320 x 320 pixel array. From the panels in Fig. 4 it is evident how the VIPA image and lineout spectrum progress as the gas fills the cell. In under 2 seconds the VIPA records the cell from vacuum to 2 mTorr with the ability to see changes in absorption from single consecutive frames [23]. For the full video demonstration see the Supplemental information section online.

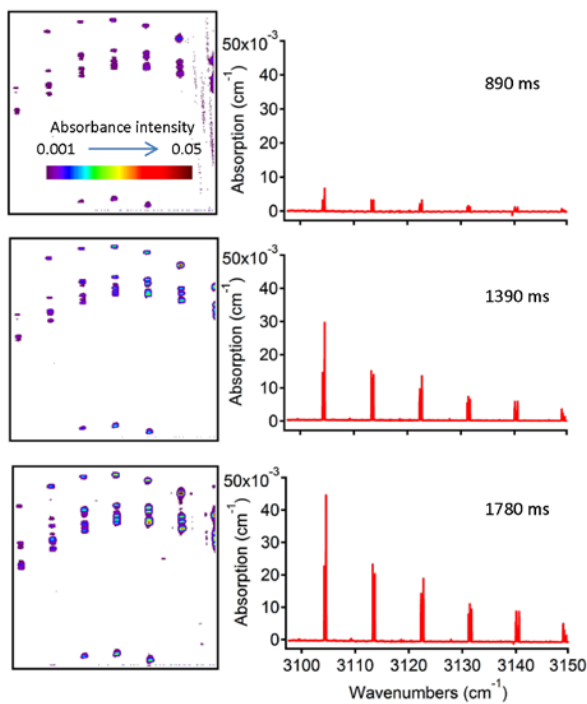

**Figure 4:** Side by side VIPA images and lineout spectra derived from those images at three different time stamps during a 375 frame/s video taken as pure methane fills a 20 cm sample cell from vacuum to ~2 Torr . A real time video can be seen in the supplemental information online.

## 3. Results
### 3.1 Linearity and dynamic range
While broadband mid-IR spectroscopy efforts have demonstrated the ability to determine the presence of a trace gas species [17, 18, 28], it is also important to quantify the amount present and the uncertainty with which the concentration can be discerned. To investigate the measurement uncertainty and linearity of this MIR spectroscopic system, we filled a Herriott cell (26 m path length, 14 passes) with static pure methane and compared the results to a pressure reading from a traditional convectron gauge (with a reported uncertainty of 0.1 mTorr at pressures below 10 mTorr). Figure 5 shows the results of this test. The top panel focuses on a single set of absorption lines at three different pressures, showing the increase in absorption with an increase in pressure. The bottom panel plots the pressure obtained by fitting the absorption data with a HITRAN [2] model (left axis) vs. the pressure reading from a traditional pressure gauge. The relationship between the fitted absorption spectrum and the pressure gauge should ideally be linear with a slope of 1, and indeed at very low pressures (up to about 4 mTorr for this path length) this is the case. The uncertainty in the fitting of a single group of methane lines (as seen in Fig. 5, top panel) was found to be on the order of 0.1%, while the methane concentration derived from the individual fits varied by 5-7% when comparing all line groups from a single image. The latter are shown as the error bars in Fig. 5.

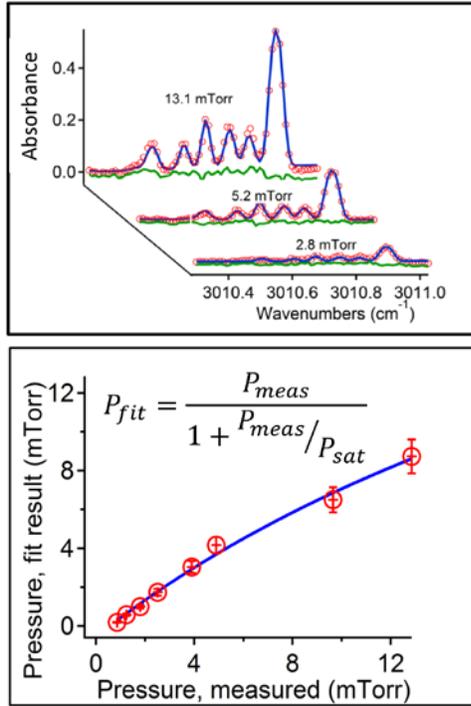

**Figure 5: Top panel:** The P(0) line group of the $\nu_3$ rovibrational transition in $CH_4$. The red dots are the data extracted from the VIPA spectrometer image, the blue line is a fit to known HITRAN values, and the green line is the fit residual. The same line grouping is shown at three different pressures in the multipass cell. **Bottom Panel:** A plot of pressure values extracted from the HITRAN fits of a 40 nm span of $CH_4$ absorption lines versus the measured pressure with a traditional pressure gauge.

As the pressure increases within the given path length, the measured absorption for some lines with large absorption cross-sections appear to saturate, causing the fitted pressure to be inaccurate. While we call this saturation of the spectral feature, this effect actually arises from the limited linearity and dynamic range of the InSb array camera, particularly at low light levels. For example, while weaker absorption lines (<0.01 peak absorption) remain within the camera's linear detection range, the stronger lines (example: >1 peak absorption) reduce the light level to a point that the camera cannot linearly detect the additional reduction in the transmission (i.e. when very few photons make it to the detector at the peak of very strongly absorbing lines). For a given path length, concentration, and incoming beam photon count, this absorption feature saturation cannot be avoided even by changing the integration time of the detector. The bottom panel of Figure 5 shows that this effect can be fit as a linear relationship including a saturation level, as in Equation 2:

$$P_{meas} = \frac{P_{gauge}}{1 + P_{gauge}/P_{sat}} \qquad \text{Equation 2}$$

Where $P_{gauge}$ is the gauge measured pressure (x axis), $P_{meas}$ is the pressure determined from the data (y axis) and $P_{sat}$ is a floating variable indicating the pressure at which full saturation occurs for this system. When $P_{gauge} \ll P_{sat}$, the equation reduces to $P_{meas}=P_{gauge}$, indicating the expected relationship when in the linear regime of the detector. However, at higher pressures, several strongly absorbing lines appear saturated as described above. This effect will become important when discussing the atmospheric data in a later section. When $P_{gauge} \gg P_{sat}$, equation 2 reduces to $P_{meas} = P_{sat}$, where the saturation value dominates. In the series of data points in Fig. 5, the fit to the data yields $P_{sat}$ = 14.5+/-0.3 mTorr. The lowest detected methane pressure of Fig. 5 was fit to give a value of 0.134 mTorr, a value that translates to ~200 ppb in a typical atmospheric pressure of our lab. In a previous paper, we reported an absorption sensitivity of 4.5 x10$^{-4}$ absorption units (in the 20 cm cell at a pressure of 2 Torr) for the laser and spectrometer system [23]. For CH$_4$ line strengths in this region, a change in absorbance of 4.5 x10$^{-4}$ corresponds to a pressure change of ~0.002 mTorr. This corresponds to a 2-3% uncertainty on the pressure value at the lowest detectable concentrations of CH$_4$, and is only slightly lower than the uncertainty returned from fit shown in Fig. 5.

### 3.2 Open air indoor path

While the data of the previous section demonstrate the measurement of relevant concentrations in a static cell, many more factors come into play when detecting similar concentrations under atmospheric conditions. To demonstrate the ability to see millitorr concentration species in atmospheric conditions, we replaced the Herriott cell with an open-air indoor path in the lab. As Figure 2 shows, we replaced the "sample" in Fig. 2(a) with a mirror to direct the beam across the lab. A beam expanding telescope was used to increase the beam size to ~3 cm in diameter in order to reduce the divergence of the beam at the 3.25 μm wavelength. After steering the beam to the retro-reflection mirror, the return beam was captured by the 25 cm diameter aperture of the telescope, resulting in a total indoor, open air path length of ~26 m. The telescope served to focus the large beam outside the back aperture. A collimating lens and two mirrors were then used to launch the light into a ~30 cm piece of MIR PCF fiber. The output of the PCF fiber serves as the entrance to the 2-D VIPA/grating spectrometer (see Fig. 2(a)). The laser spectrum, when optimized at one non-linear crystal position in the OPO by tuning the pulse width and pump power, covers up to 300 nm of bandwidth. Therefore it is necessary to rotate the grating in the spectrometer slightly, using 3-4 images to cover the entire laser bandwidth. Each image at this low light level requires 2 ms camera acquisition time. One dark image was taken first (by blocking the laser beam), and then for each grating position a signal (beam going out to the retro-reflection mirror) and background (beam traversing a short path to the telescope, see Fig. 2(c)) image was taken.

Fig. 6 contains the results of analyzing a single VIPA image (as explained in Fig. 3) and converting this image to a traditional line out spectrum for a long path indoor measurement. The data, shown in red, was fit with a modeling program we have developed that is based on HITRAN data for line positions, and temperature and pressure broadening, as well as a background correction [2]. After a general frequency scale is established by calibrating the raw absorbance peaks with a simulated spectrum from peaks in the HITRAN data base for H$_2$O and CH$_4$, the fitting program is given the absorption data, a calibrated frequency axis, a fixed atmospheric pressure value (which governs the pressure broadening of the peaks), and a fixed instrument response value [23]. Our fitting program minimizes the residuals (difference between the absorption data and the HITRAN model) by iterating the concentration of H$_2$0 and CH$_4$, the self-pressure broadening (given by the partial pressure of the gases themselves) given by HITRAN, and allows for a linear frequency correction across the spectrum.

In a 26 m path length, several of the measured water absorption peaks appear saturated (due to detector linearity at low light levels as described above in section 3.1). This saturation level was determined in the analysis of Fig. 5. In some cases, the water peak absorption values can be 10-20x larger than the methane lines, largely due to the ~4000:1 concentration ratio of water to methane in the atmosphere. In order to clearly see the methane peaks above the dark count noise of the detector, the acquisition time is set at 2 ms. Due to the line strength discontinuity between water and methane lines, we have cut the absorption spectrum at an absorption value of ~0.7, corresponding to 50% transmission. This cutoff level was determined by first fitting the spectrum without cutting off the saturated peaks, and using this residual to determine the saturation level. Without removing the tops of the saturated peaks, the residuals reach from 0.4 to -0.4 absorption units as the fitting program attempts to "average" out the concentration of the saturated water peaks with the non-saturated peaks. By removing the tops of the saturated peaks, we are able to reduce the residuals by a factor of 3.5. This allowed us to determine the water concentration more accurately, as the saturation only affects the spectral shape of the top of the water absorption peaks. The methane peaks are all below this saturation limit/cutoff. In the top panel of Fig. 7, the actual data is shown in a dotted red line, the cut spectrum used for fitting is shown in a solid red line, and the fit is shown as its negative in blue. The residuals (in grey, above) have some small spikes, which we attribute to errors in the frequency scale across the 100 cm$^{-1}$ bandwidth of a single image. The true center of the saturated peak is difficult to determine when fitting the data set as a whole. Nevertheless, the residuals have an rms value of 0.021 absorbance units over the 100 cm$^{-1}$ bandwidth with peaks ranging in magnitude from 0.2 to 4 absorbance units.

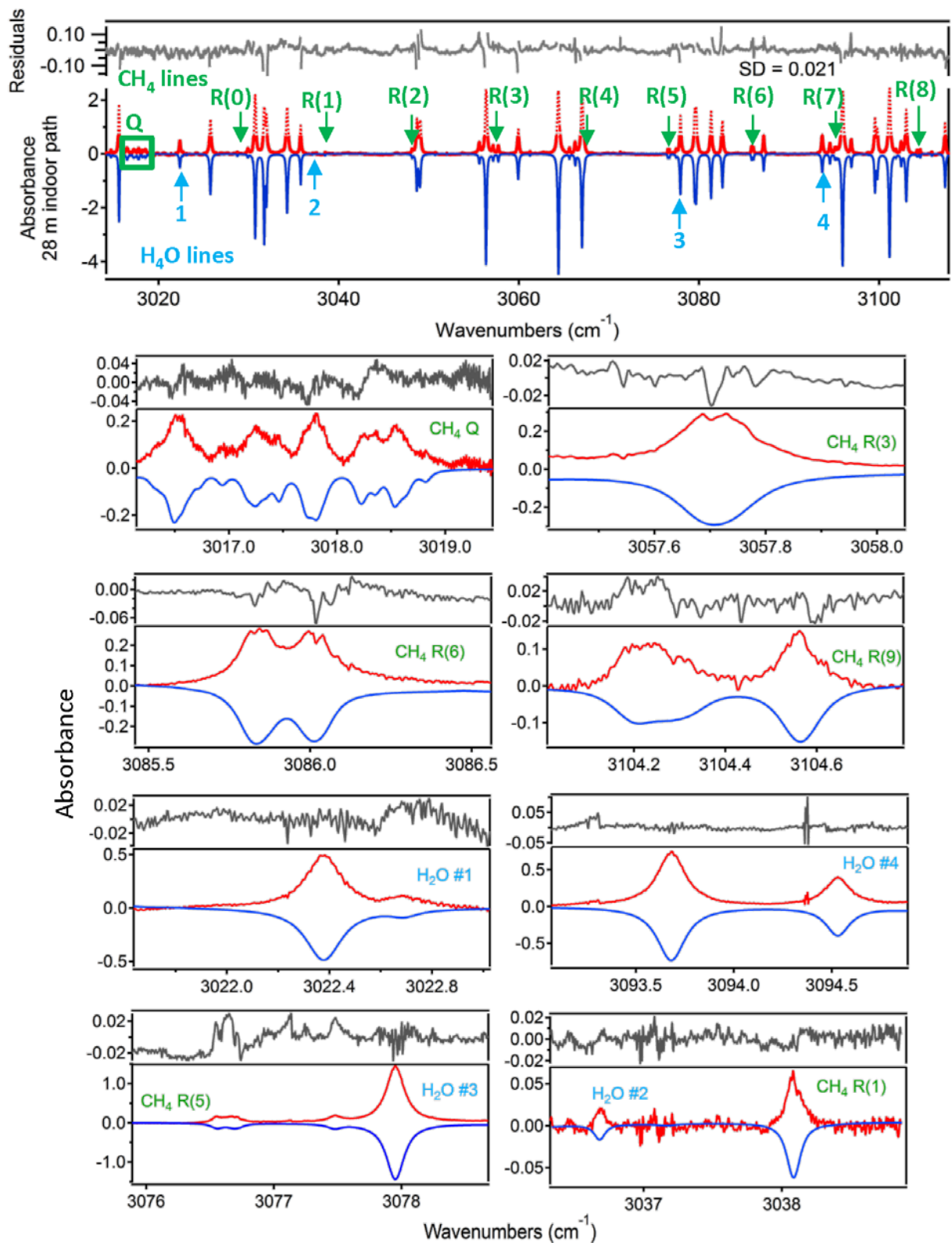

**Figure 6:** Data from a single VIPA image (dotted red curve) for an open-air indoor measurement (26 m). The spectrum was cut at 50% transmission before fitting due to absorption line saturation as discussed in the text (solid red line). The blue curve, displayed as its negative for clarity, is a global fit of the data with a HITRAN model described in the text. The residuals of the fit are shown in grey above the data. The bottom eight panels are zoom-in fits of isolated methane and water lines and their residuals.

One of the difficulties of obtaining open air data is the background subtraction. Even with best efforts to ensure that the background image and long path image are identical except for the sampling path length, systematic offsets in the absorption baseline still exist and removing these offsets correctly is the current limit to the sensitivity of this measurement. This will be discussed further in the conclusions section. For comparison, we have developed an alternative approach in which we cut out portions of the spectrum and fit them individually after removing the background in the vicinity of specific spectral features (Figure 7). This background subtraction method requires only a single data image, eliminating the need for background images. We created a filter using a model HITRAN spectrum to indicate where the absorption should be within 1% of zero (see Figure 7(a)). Then we fit the filtered raw signal data (excluding the peaks) so that the "background" was a fit to the zero absorption portions of the spectrum (Fig. 7(b)). After removing this background from the measured absorption, each individual line group was fit and concentrations were determined from that fit (Fig. 7(c)). Selected peaks can be seen in Figure 6, where individual fits of line groups for both methane and water are displayed. These individual results are summarized in Fig. 8 (top panel: methane, bottom panel: water). The data points (circles) represent the value of methane or water from individual line fits. While the values are scattered, the average of the individual fits (top panel: red, CH$_4$ average, bottom panel: blue, H$_2$O average) is in good agreement with the global fit for the both the methane and the water values. The shaded box is the standard deviation of the individual fit data. The methane concentration determined from the global fit (obtained using dark, background and sample images) was 1.96 ppm (dashed green line in Fig. 8), and the average of individual line fits (using a single sample image) was 1.92 +/- 0.13 ppm (dashed red line in Fig. 8). In this instance we did not have an accurate second measurement for comparison of the concentration. However, the fitted values are in good agreement with typical concentration values of methane in the air, which vary from 1.8-2.0 ppm, and the global and individual fits were self-consistent. The water concentration was determined to be 1.2% in both the global fit and the average of the individual fits, with a 0.1% standard deviation on the average (show as the shaded box in Fig. 8). The water concentration can vary daily from 0.5-2%, and our fitted value is within this range. For comparison, without cutting the absorption spectrum to account for saturated peaks, the water value was determined to be 0.6% in the global fit.

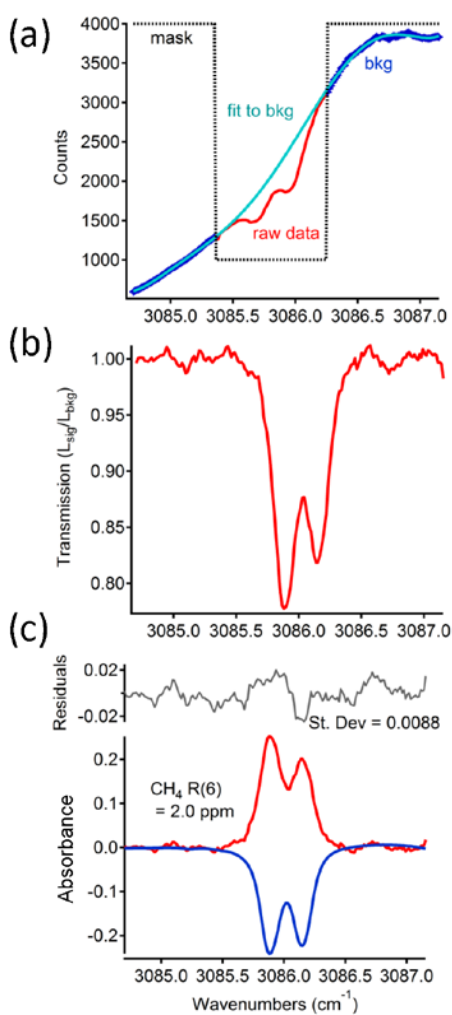

**Figure 7: (a)** Local lineout data around an isolated methane line from a single raw data image (red). A mask created from a simulated HITRAN spectra determines where the %T value should be near zero. The zero-line points are fit with a polynomial curve of order 5. This curve then becomes the "background". Then the true %T signal **(b)** is calculated as data/background. When the corrected transmission data is fit with a HITRAN model, the absorption spectrum and resulting residuals are shown in **(c)**. The standard deviation of the fit residuals is ~2.5% of the peak absorption.

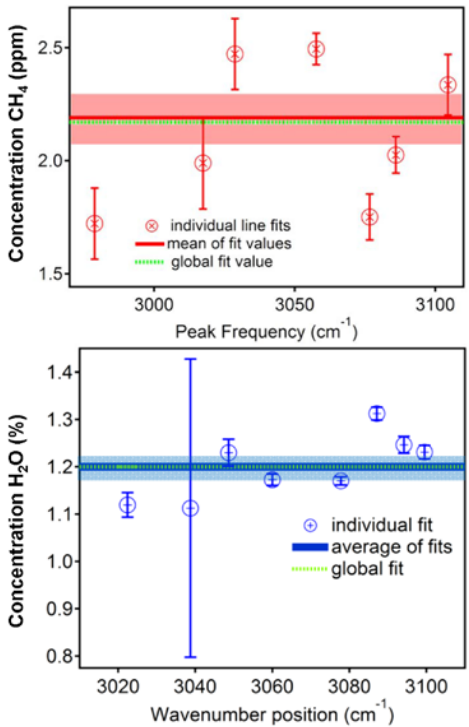

**Figure 8: Top panel:** Methane concentration determined from the data in Fig. 5. The green dotted line is the concentration obtained from the global fit of ~100 cm$^{-1}$ spectrum. The red circles are from individual fits with individual background subtractions around that line group. The red line indicates the average of the individual fits (2.11 +/- 0.13 ppm) with an error represented by the red shaded rectangle. **Bottom Panel:** Similar to the top panel except for water concentration. The green dotted line is again the global fit, and the blue circles and line are the individual fits and average (1.2 +/-0.1%), respectively.

### 3.2 Open-air outdoor path

While Fig. 6, 7 and 8 represent indoor open-air measurements, an outdoor open-air measurement has several other considerations. Turbulence, precipitation, temperature and pressure variations, and etalons from exit windows come into play when making an outdoor measurement. To further test our ability to monitor trace gas concentrations in an outdoor environment, we moved the laser and OPO system into a lab where a window to the outside was accessible. Fig. 2 (d) shows the window access with the laser system on its 120 cm by 120 cm table positioned adjacent to the window. The beam was directed out and retro-reflected back through a MIR transparent 25 cm diameter window. The beam traversed grassy terrain and the retro-reflecting mirror was placed on a small tripod platform on the grass (not pictured). A path length of ~25 m was used for the outdoor measurements. These results are shown in Fig. 9(a). The data is inherently noisier due to the majority of the path length occurring outdoors, however, a global fit of the data is still possible and gives a methane value of 2.17 ppm. Fig. 9(b) zooms in on the Q branch of the CH$_4$ band from the global fit. The residuals show the background variation underlying the data. Fig. 9(c) uses the raw data from a single VIPA image as described earlier, and a slightly cleaner fit is obtained. During the outdoor measurement an independent methane measurement was available [7]. The independent measurement used an intake air tube that was positioned near the path of the MIR beam. The outside air was brought into a mobile device where a single frequency laser was used to determine the methane concentration, using a chemically produced "zero methane" air sample as a background. In this way we were able to compare our measured value to an independent value. Figure 9(d) summarizes the methane results. The dotted green line represents the methane concentration value for a global fit (2.17 ppm) of the data in Fig. 9(a). As with the indoor open path data, methane lines are taken individually and fit using the method described above (with no background image), and the results are shown as circles in Fig. 9(d), with the average of these points displayed as the solid red line (2.19 +/-0.13 ppm). The independent measurement of methane was 1.94 ppm, which is slightly lower than both the global fit and the average of individual fits of methane lines. At this point, we attribute this to discrepancy to small systematic offsets due to troublesome background subtraction and noisier data, and possibly to part of the path length being inside. Other differences in the two measurement techniques were that the intake air tube was 150 cm off the ground in a single position, and the 25 m path length of our laser traversed grassy terrain, ranging from 150 cm to 30 cm above the ground. Figure 9(e) shows the same data but for the water concentration (also measured independantly with the methods of ref [7] and other humidity measurement stations in the area). The global value is 0.56% when using a spectrum cut at 50% transmission to account for saturated peaks (as described in section 3.1) . The individual fit values, given by the blue circles in Fig. 9(e) give an average of 0.57 +/- 0.04 %. Again, both values from the global fit and average fits are larger than an independant measurements of 0.4 %, but are consistent with relative humidity data values of 0.60% and 0.52% from two monitoring stations in Boulder. The residuals of the global fit (spanning 280 cm$^{-1}$ of bandwidth from 4 concatenated VIPA images) give a standard deviation of 0.024 absorbance units.

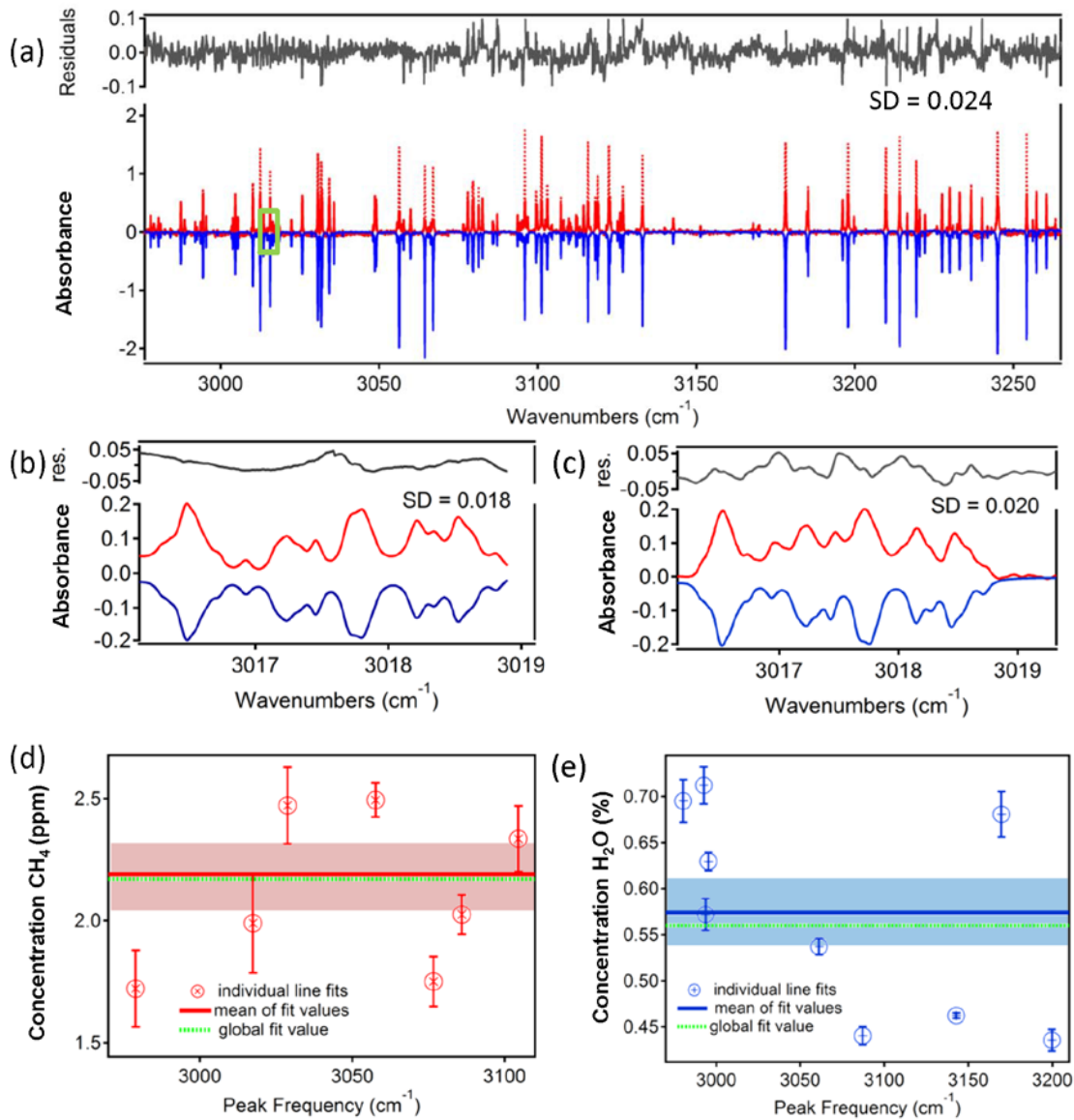

**Figure 9: (a)** A nearly 250 cm-1 span of a 26 m path of outdoor air, including four images concatenated together. Each image had an acquisition time of 2 ms. The red curve is the data, the blue curve is the fit based on HITRAN methods described in the text, shown in its negative for clarity, and the gray curve is the residual of the fit. **(b)** A zoom in of the global fit in the Q branch of the methane $\nu_3$ band (the green box in (a)) **(c)** A fit of the same Q branch data using a background subtraction from the raw image and using only the background in the immediate area around the peaks. **(d)** Methane individual line fits (average 2.19 +/- 1.3 ppm, shown as solid red line) compared with the global fit value from (a) of 2.17 ppm, shown as dotted green line. Independent measurement sources yielded a value of 1.94 ppm. **(e)** Water individual line fits (average 0.57 +/- 0.04 %, shown as solid blue line) compared to the global fit value of 0.56%, shown as solid green line. Independent measurement sources gave values of 0.41 %, 0.60 % and 0.51 %.

## 4. Conclusions

To our knowledge, this work is the first demonstration of an outdoor open air measurement at 3.25 μm utilizing a 2-D VIPA spectrometer. We have demonstrated the use of a broad bandwidth mid-infrared laser system to detect small concentrations of isolated species with fast acquisition times (100's of μs to 2 ms). In a lab setting, the background subtraction yields clean absorption data and concentrations of pure methane were measured down to ~200 ppb with an uncertainty of 27 ppb (13.6%, taken from the residuals of the fit to the data at the lowest pressure measurement). In static cell situations, 50 nm of spectral bandwidth data can be acquired with a rate of 375 frames/s (using a reduced 320 pixel x 320 pixel image array), which has applications for the monitoring of a rapidly changing species concentration. The accuracy of static cell measurements utilizing the fitting of direct absorption data were found to be in good agreement with traditional pressure measurements.

The system described here, due to the brightness and spatial coherence of the MIR light at 3.25 μm, is also useful for broadband detection of multiple species in an open-air outdoor path. We demonstrate the detection of small concentrations of $CH_4$ in the atmosphere (~2 ppm) in the presence of much higher absorption water lines. From unsaturated individual lines, we were also able to determine the water concentration over a range of 0.5-1.2%. Both the methane and water concentrations are consistent with independent data measurements. At the present time, the determination of the absorption baseline is the limiting factor in the measurement of the true concentration of a species in the open air. Many factors affect the background, including systematic etalons from exit windows, pointing stability of the laser beam into the fiber, laser amplitude noise or drift, as well other environmental factors. Both methods of background subtraction have difficulties. Comparing two images taken seconds apart requires more time and effort to minimize differences between images. Using a piece-wise background fitting of a raw data image, as we have shown, is most effective if there are enough zero transmission points around the peaks of interest. We report measurement uncertainties on the measured concentrations on the order of ~6-8% in the open air data (both indoor and outdoor, for both water and methane). We use two separate methods to determine the concentration and the values are internally consistent. The uncertainty is significantly reduced down to 2% when making static cell measurements. For comparison, Ref. [7] reports that an international standard for intra-laboratory comparability measurements for methane is +/-2 ppb, or ~0.1% of the atmospheric concentration of methane. While our uncertainties are larger than the state-of-the-art, they still have the advantage of requiring only a 2 ms integration time for a 100 nm spectral bandwidth in the mid-IR region, covering multiple species in a single image.

The authors thank Tyler Neely and Florian Adler for assistance in building and testing the Yb laser, amplifier, OPO and VIPA spectrometer, and Fetah Benabib for providing the air–core PCF. We are also grateful to J. Kofler, G. Petron, C. Sweeney, and P. Tans from the National Oceanic and Atmospheric Administration in Boulder, CO for the independent measurement of methane and water during the outdoor measurements. We are indebted to Nathan Newbury for sharing his lab space, equipment and expertise as well as Flavio Cruz for helpful discussions. This work is supported by NIST and is a contribution of the US government; it is not subject to copyright in the US.

**References**

1. I. Levin, T. Naeglar,, B. Kromer, M. Diehl, R. J. Francey, A. J. Gomez-Pelaez, L. P. Steele, D. Wagenbach, R. Weller and D. E. Worthy, Tellus Series B: Chemical and Physical Meterology **62B**: 26 (2010)
2. L.S. Rothman, et.al, Journal of Quantitative Spectroscopy and Radiative Transfer **110**, 533 (2009)
3. P. Weibring, D. Richter, J. Walega, A. Fried, Optics Express **15** (21), (2007)
4. K.C. Cossel, F. Adler, K. A. Bertness, M. J. Thorpe , J. Feng, M. W. Raynor, and J. Ye, Applied Physics B **100,** 917 (2010)
5. M. Thorpe, D. Clausen, M. Kirchner, and J. Ye, Optics Express **16**(4), 2387 (2008)
6. G. Rieker, F. Giorgetta, W. Swann, J. Kofler, A. Zolot, L. Sinclair, E. Baumann, C. Cromer, G. Petron, C. Sweeney, P. Tans, I. Coddington, and N. R. Newbury, arXiv:1406.3326 (2014).
7. C. W. Rella, H. Chen, A. E. Andrews, A. Filges, C. Gerbig, J. Hatakka, A. Karion, N. L. Miles, S. J. Richardson, M. Steinbacher, C. Sweeney, B. Wastine, and C. Zellweger, Atmospheric Measuremnt Techniques **6**, 837 (2013)
8. A. Fleisher, B. Bjork, T. Bui, K. Cossel, M. Okumura, and J. Ye, Journal of Physical Chemistry Letters **5**(13), 2241 (2014)
9. I. Galli, S. Bartelini, S. Borri, P. Cancio, D. Mazzotti, P. De Natale, and G. Giusfredi, Physical Review Letters **107**, 270802 (2011)
10. R. Curl, F. Capasso, C. Gmachl, A. Kosterev, B. McManus, R. Lewicki, M. Pusharsky, G.Wysocki, F. Tittel, Chemical Physics Letters **487**, 1 (2010)
11. P.Jouy, M. Mangold, B. Tuzson, L. Emmenegger, Y. Chang, L. Hvozdara, H. P. Herzig, P. Wägli, A. Homsy, N. deRooij, A. Wirthmueller, D. Hofstetter, H.t Looserg and J. Faist, Anaylst **139** (9), 2039 (2014)
12. C. Xia, M. Kumar, M. Cheng, R. Hegde, M. Islam, A. Galvanauskas, H. Winful, and F. Terry, Jr., Optics Express **15** (3), 865 (2007)
13. B. Bernhardt, E. Sorokin, P. Jacquet, R. Thon,T. Becker, I. T. Sorokina, N. Picqué, T.W. Hänsch, Applied Physics B **100**, 3 (2010)
14. Ł. Kornaszewski, N. Gayraud, J. M. Stone, W. N. MacPherson, A. K. George, J. C. Knight, D. P. Hand, and D. T. Reid, Optics Express **15**(18) (2007)
15. F. Adler, K. Cossel, M. Thorpe, I. Hartl, M. Fermann, and J. Ye Optics Letters **34**(9), 1330 (2009)
16. A. Hugi, G. Villares, S. Blaser, H. C. Liu, and J. Faist, Nature **492**, 229 (2012)
17. A. Schliesser, N.Picque, T. W. Hansch, Nature Photonics **6**, 440 (2012)
18. F. Adler, P. Maslowski, A. Foltynowicz, K. C. Cossel, T. C. Briles, I. Hartl, and J. Ye, Optics Express, **18**(12), 21861 (2010)
19. A. M. Zolot, F. R. Giorgeta, E. Baumann, J. W. Nicholson, W. C. Swann, I. Coddington, and N. R. Newbury, Optics Letters **37**(4), 638 (2012)
20. E. Baumann, F. R. Giorgetta, W. C. Swann, A. M. Zolot, I. Coddington, and N. R. Newbury, Physical Review A **84**, 062513 (2011)
21. T. Ideguchi, A. Poisson, G. Guelachvili, N. Picque, and T. W. Hansch,Nature Communications, **5** (3375) (2014)
22. T. Johnson, S. Diddams, Applied Physics B **107**(1), 31 (2012)
23. L. Nugent Glandorf, T. Neely, F. Adler, A. Fleisher, K. Cossel, B. Bjork, T. Dinneen, J. Ye, and S. Diddams, Optics Letters **37**(15), 3285 (2012)
24. L Nugent-Glandorf, T. Johnson, Y. Kobayashi, and S. A. Diddams*,* Optics Letters, **36**(9), 1578 (2001)

25. M. Shirasaki, FUJITSU Sci. Tech. Journal, **35**(1), 113 (1999)
26. F. Benabid, Philosophical Transactions of the Royal Society A, 364, 3439 (2006)
27. S. Xiao, A. Weiner, and C. Lin, IEEE: Jounal of Quantum Electronics **40**(4), 420 (2004)
28. K. L. Vodopyanov, E. Sorokin, I. T. Sorokina, and P. G. Schunemann Optics Letters **36**(12), 2275 (2011)